\documentclass[preprint,showpacs,showkeys]{revtex4}

\usepackage{amsmath}
\usepackage[pdftex]{graphicx}
\DeclareGraphicsExtensions{.pdf,.png,.jpg}
\usepackage{subfig}
\usepackage{bm}
\usepackage{color}

\begin{document}

\title[Calibrating spectral estimation for the LISA Technology Package with
multichannel synthetic noise generation]{Calibrating spectral estimation for the
LISA Technology Package with multichannel synthetic noise generation}

\author{Luigi Ferraioli}
\email{luigi@science.unitn.it}
\affiliation{University of Trento and INFN, via Sommarive 14, 38123 Povo (Trento), Italy}

\author{Gerhard Heinzel}
\affiliation{Albert-Einstein-Institut, Max-Planck-Institut fuer Gravitationsphysik und Universitaet Hannover, Callinstr. 38, 30167 Hannover, Germany}

\author{Martin Hewitson}
\affiliation{Albert-Einstein-Institut, Max-Planck-Institut fuer Gravitationsphysik und Universitaet Hannover, Callinstr. 38, 30167 Hannover, Germany}

\author{Mauro Hueller}
\affiliation{University of Trento and INFN, via Sommarive 14, 38123 Povo (Trento), Italy}

\author{Anneke Monsky}
\affiliation{Albert-Einstein-Institut, Max-Planck-Institut fuer Gravitationsphysik und Universitaet Hannover, Callinstr. 38, 30167 Hannover, Germany}

\author{Miquel Nofrarias}
\affiliation{Albert-Einstein-Institut, Max-Planck-Institut fuer Gravitationsphysik und Universitaet Hannover, Callinstr. 38, 30167 Hannover, Germany}

\author{Stefano Vitale}
\affiliation{University of Trento and INFN, via Sommarive 14, 38123 Povo (Trento), Italy}

\begin{abstract}

The scientific objectives of the Lisa Technology Package (LTP) experiment on board
of the LISA Pathfinder mission demand for accurate calibration and validation of the
data analysis tools in advance of the mission launch. The level of confidence
required in the mission outcomes can be reached only by intensive testing the tools
on synthetically generated data. A flexible procedure allowing the generation of
cross-correlated stationary noise time series was set-up. Multichannel time series
with the desired cross-correlation behavior can be generated once a model for a
multichannel cross-spectral matrix is provided. The core of the procedure comprises
a noise coloring, multichannel filter designed via a frequency-by-frequency
eigen-decomposition of the model cross-spectral matrix and a subsequent fit in the
Z-domain. The common problem of initial transients in filtered time series is solved
with a proper initialization of the filter recursion equations. The noise generator
performance was tested in a two-dimensional case study of the closed-loop LTP
dynamics along the two principal degrees-of-freedom.

\end{abstract}

\pacs{04.80.Nn, 95.75.-z, 07.05.Kf}

\keywords{spectral estimation; calibration; LISA Pathfinder; LISA Technology Package; LTP; synthetic noise generation; multichannel systems}

\maketitle

\section*{Introduction}

The LTP (LISA Technology Package) experiment is the main scientific payload on the
European Space Agency mission, LISA Pathfinder. Its goal is to determine and analyse
all possible sources of disturbance which perturb the free-falling test masses
from their geodesic motion. The system is composed of two test masses whose
position is sensed by an interferometer. The spacecraft cannot simultaneously
follow both masses, and so the trajectory of only one test mass serves as the
drag-free reference along the $x$ (measurement) axis. To prevent the trajectories
of the two masses from diverging in response to any differential force, the
second test mass is electrostatically actuated to follow the spacecraft. In the main
science operating mode, the position of the SC relative to the first test mass
is controlled using micro-Newton thrusters attached to the SC. The position of the
second test mass is controlled using capacitive actuators surrounding the test
mass. The first interferometer channel measures the position of the first test mass
relative to the spacecraft. The second interferometer channel (differential channel)
measures the relative displacement between the two test masses.

A set of different experiments, such as measurements of parasitic voltages, test
mass charging, thermal and magnetic disturbances, completely covers the scheduled 90
days of LTP operations; the overall aim of the experiments is to reach the best
free-fall quality in a step by step procedure in which the result of the previous
experiment is used to define the detailed configuration of the following experiment.
This cascade-like process aims to demonstrate the ability to put a test mass into
free-fall at a level where any residual acceleration is below $3 \times 10^{-14}
\textnormal{ m} \textnormal{ s}^{-2} / \sqrt{\textnormal{Hz}}$ at frequencies around
$1$ mHz \cite{Armano, Anza, McNamara, Bortoluzzi1, Bortoluzzi2}.

Such a demanding accuracy requires a careful calibration of the spectral estimation
algorithms so as to avoid any systematic bias in the estimation of the spectrum of
the residual acceleration. Due to the limited time duration of the mission, the
amount of data available will be not enough for a meaningful and robust calibration
of the dedicated data analysis tools. The natural way to solve the problem is to
calibrate and test the tools in advance of the mission, by an in-depth analysis of
synthetic noise data. The experiment has a total of $18$ measuring channels sensing the movement of the test masses, many of
which are coupled so that information is contained not only in the individual
power-spectra, but also in the cross spectral densities between different channels.
In order to set-up a reliable test bench for such a system, a robust and flexible
multichannel noise generator is required.

The problem of generating a sequence of random variables having some definite
statistical properties is well examined in literature. Stein and Storer \cite{stein}
proposed a procedure for which the computation of $N$ sample values requires the
eigen-decomposition of the covariance matrix of the process. This allows the
identification of a transformation matrix that multiplies a vector of independent
samples to provide a noise series with the desired correlation properties. Levin
\cite{Levin} suggested instead to pass white noise through a digital noise coloring
filter with a rational transfer function. The problem connected with the initial
transient is solved with the calculation of $K$ consecutive ($K$ is the order of the
noise shaping filter) output values having the same statistical properties as if
they were produced by steady-state operation of the filter. An alternative method
for filter initialization was indicated by Kay \cite{Kaypaper} who realized that one
has just to specify the initial state for the filter. The method for the calculation
of the initial state is based on the Levinson-Durbin algorithm. As an alternative,
Franklin \cite{Franklin} described a procedure for the simulation of stationary and
non-stationary Gaussian random processes. The procedure for a non-stationary process
is similar to that reported in \cite{stein} and is based on the Crout factorization
of the covariance matrix of the process. The output for the stationary case is,
instead, the result of a simulation of a continuous system by means of the state
space formalism. The initial state is calculated with a linear transformation from
uncorrelated random noise samples. All the methods in literature deal with the
generation of a single channel of data with a given correlation function or,
analogously, with a given spectrum. The algorithms proposed by Levin, Kay and
Franklin can be crudely summarized in three steps; 1) identification of the desired
system, 2) initialization of the data sequence, 3) generation of the colored noise
data sequence from a sequence of zero mean, delta correlated random numbers.

The method proposed in the present paper follows this classical scheme with the
relevant difference that it is designed to work with \emph{multichannel} systems
(i.e. multiple inputs, multiple outputs). In the following, the mathematical basis
of the method is developed, and a case study is discussed in order to quantitatively
assess the reliability of the procedure.

\section{Principles of the noise generation procedure}\label{sec:principles}

It can be assumed, in complete generality, that the noise to be generated
$x\left(t\right)$ has a power spectral density (PSD) that can be written as:

\begin{equation}
	S_{xx}\left(\omega\right) = \left|H\left(\omega\right)\right|^2 S_0.
	\label{eqn:Sxx=HS0}
\end{equation}

The process $x\left(t\right)$ can be thought as the output of a rational continuous
filter with transfer function $H\left(\omega\right)$, at the input of which is a
white, zero-mean noise $\epsilon\left(t\right)$ with PSD equal to $S_0$. The filter
$H\left(\omega\right)$ can be written as:

\begin{equation}
	H\left(\omega\right) = \sum_{h=1}^{N} {\frac{r_h}{\imath\omega - p_h}} ,
	\label{eqn:Hpfs}
\end{equation}

with $r_h$ the residue of $H\left(\omega\right)$ in $p_h$ \footnote{Laplace notation
is adopted in the definition of the partial fraction expansion of the transfer
function $H\left(\omega\right)$: $\imath\omega \rightarrow s \Rightarrow
H\left(\omega\right) \rightarrow H\left(s\right)$. The convention adopted for the
Fourier transform is that matching Laplace transform: $H\left(\omega\right) =
\int\limits_{-\infty}^\infty h\left(t\right)e^{-\imath \omega t} \, dt$ and
$h\left(t\right) = \frac{1}{2 \pi} \int\limits_{-\infty}^\infty
H\left(\omega\right)e^{\imath \omega t} \, d\omega$. Since $p_h$ are simple poles,
the calculation of residues can be performed with the rule
$Res\left({H\left(s\right),p_h}\right) = \lim_{s \to p_h}
\left(s-p_h\right)H\left(s\right)$. }.

Equation (\ref{eqn:Hpfs}) shows that the process $x\left(t\right)$ can in turn
be considered as:

\begin{equation}
	x\left(t\right) = \sum_{h=1}^{N}{y_h\left(t\right)},
	\label{eqn:x=yht}
\end{equation}

where:

\begin{equation}
	y_h\left(t\right) = r_h \int\limits_0^\infty e^{p_h t'} \epsilon\left(t-t'\right) \, dt'.
\end{equation}

Thus, generating the process $x\left(t\right)$ is equivalent to generating the $N$
correlated processes $y_h\left(t\right)$. A discrete process with time-step $T$ can
be realized from the recursive equations:

\begin{eqnarray}
	y_h\left(t+T\right) & = & y_h\left(t\right) e^{p_h T} + \epsilon_h\left(t+T\right) \nonumber \\
	\epsilon_h\left(t+T\right) & = & r_h \int\limits_0^T e^{p_h t'} \epsilon\left(t+T-t'\right) \, dt' \nonumber \\
	h & = & 1, \ldots, N.
\end{eqnarray}

Since the procedure must not diverge, only poles $p_h$ with a negative real part
(stable poles) are considered. The processes $\epsilon_h\left(t+T\right)$ are not
independent but it can be verified that their cross-correlations is vanishing for
time intervals larger than $T$. Then, indicating with $k$ and $m$ integers values,
the cross-correlation between input processes can be written as:

\begin{equation}
 \langle \epsilon_i\left(k T\right) \epsilon_j\left(m T \right) \rangle  = \delta_{k m} \frac{S_0 r_i r_j}{p_i + p_j} \left[e^{\left(p_i + p_j\right) T} - 1\right],
	\label{eqn:ccorreps}
\end{equation}

where the notation $\left\langle \right\rangle$ indicates the expectation value
operator. If the process $\epsilon\left(t\right)$ is zero-mean Gaussian, then so is
$\epsilon_j \left(k T\right)$, and equation (\ref{eqn:ccorreps}) is sufficient to
determine the statistics.

The generated series $x\left(k T\right)$ exactly represents the sampling of the
continuous process $x\left(t\right)$ with a time-step $T$, and therefore it also
reproduces the aliasing if the filter $H\left(\omega\right)$ has a response
different from zero at frequencies larger than $1/2T$. This is a consequence of the well known relation between the spectra of discrete and continuous processes \cite{Percival}:

\begin{eqnarray}
	\bar{S}_{xx}\left(\omega\right) & = & \sum_{k = -\infty}^{\infty} S_{xx}\left(\omega + \frac{2 \pi k}{T}\right) \nonumber \\
	\left|\omega\right| & \leq & \frac{2 \pi k}{T}.
	\label{eqn:disccont}
\end{eqnarray}

Here $T$ is the sampling time, $\bar{S}_{xx}\left(\omega\right)$ and $S_{xx}\left(\omega\right)$ are the spectra of the discrete and continuous processes respectively.

From the point of view of the numerical implementation, it is more convenient to
start from the assumption of a discrete filter. In the case of finite length
discrete time series, equations (\ref{eqn:Sxx=HS0}) and (\ref{eqn:Hpfs}) can be
rewritten as:

\begin{eqnarray}
	S_{xx}\left(\Omega\right) & = & \left|H\left(\Omega\right)\right|^2 S_0. \nonumber \\
	H\left(\Omega\right) & = & \sum_{h=1}^{N} {\frac{r_h}{1 - p_h e^{-\imath \Omega}}},
	\label{eqn:Hpfz}
\end{eqnarray}

where $\Omega$ is the normalized angular frequency $\Omega = 2 \pi f/f_s$, and $f_s$
is the sampling frequency.

Each element of the partial-fraction expansion in equation (\ref{eqn:Hpfz}) can be
considered as a simple autoregressive moving average filter for which the usual
recursive relation holds \cite{papoulis}:

\begin{eqnarray}
	x\left(n\right) & + & a_1 x\left(n-1\right) + \cdots + a_N x\left(n-N\right) = b_0 i\left(n\right) + \cdots + b_M i\left(n-M\right) \nonumber \\
	a_1 & = & - p    \text{ and }  a_2, \ldots, a_N = 0 \nonumber \\
	b_0 & = & r      \text{ and }  b_1, \ldots, b_M = 0.
\end{eqnarray}

Here, $a_i$ are the coefficients of the denominator polynomial, $b_j$ are the
coefficients of the numerator polynomial, $r$ and $p$ are residues and poles as
written in equation (\ref{eqn:Hpfz}) and $i\left(n-k\right)$ is the step $k$ delayed
input to the system. Thus the complete noise generation process is obtained by:

\begin{eqnarray}
	x\left(n\right) & = & \sum_{k=1}^{N} x_k \left(n\right) \nonumber \\
	x_k \left(n\right) & = & p_k x_k \left(n - 1\right) + r_k \epsilon\left(n\right).
	\label{eqn:recurrence}
\end{eqnarray}

Each $x_k\left(n\right)$ can be generated according to the recursive equation
(\ref{eqn:recurrence}) starting from the same white noise series
$\epsilon\left(n\right)$.

Such a procedure provides a noise series whose PSD is an accurate replica of the
continuous noise spectrum up to the Nyquist frequency. Aliasing is not reproduced in
the discrete case. In the rest of the paper, the detailed calculations for the
implementation of a discrete multichannel procedure are presented.

\section{Multichannel noise generation}

\subsection{Noise Coloring Filter Identification}\label{subs:noisecoloring}

A multichannel sequence can be described by the $M$-dimensional vector:

\begin{equation}
\mathbf{y}\left( t \right) = \left( {\begin{array}{*{20}c}
   {y_1 \left( t \right)}  \\
    \vdots   \\
   {y_M \left( t \right)}  \\
\end{array}} \right),
\label{multichannelprocess}
\end{equation}

where $y_i \left( t \right)$ is the data sequence at the \emph{i}th channel. If the
process is stationary, the cross-correlation matrix at a given delay $\tau$ is
defined as \cite{kay}:

\begin{equation}
\mathbf{R} \left( \tau  \right) = \int\limits_{-\infty}^{\infty} {\mathbf{y}\left( t \right)\mathbf{y}^\dagger  \left( {t + \tau } \right)} \mathrm{d}t.
\label{crosscorr}
\end{equation}

The elements of the matrix $\mathbf{R} \left( \tau \right)$ are the
cross-correlations between the different elements of the multichannel sequence. The
symbol $\dagger$ indicates a matrix conjugate transpose. The cross-spectral density
matrix for the given multichannel process is defined as the Fourier transform of
the cross-correlation matrix:

\begin{eqnarray}
\mathbf{S}\left( \omega \right) & = & \int\limits_{-\infty}^{\infty} \mathbf{R}\left( \tau  \right)\exp\left( - \imath \omega \tau \right) \mathrm{d}\tau \nonumber \\
	\mathbf{S}\left( \omega \right) & = & \left( {\begin{array}{*{20}c}
	   {S_{11} \left( \omega \right)} & \cdots & {S_{1M} \left( \omega \right)}  \\
	   \vdots                    & \ddots &                     \vdots \\
	   {S_{M1} \left( \omega \right)} & \cdots & {S_{MM} \left( \omega \right)}  \\
	\end{array}} \right).
\label{cps}
\end{eqnarray}

A noise coloring multichannel filter is a linear operation which transforms a delta
correlated unitary variance multichannel random sequence (multichannel white noise
$\bm{\varepsilon} \left( t \right)$) in to a noise sequence with the given
cross-spectral density matrix.

\begin{eqnarray}
		y_i\left( t \right) = \sum\limits_{j = 1}^N \int\limits_{-\infty}^{\infty} h_{ij}\left( \tau  \right) \varepsilon_j\left( t - \tau  \right) \mathrm{d}\tau \nonumber \\
		\left\langle \varepsilon_i\left( t \right) \varepsilon_j\left( t + \tau \right) \right\rangle = \delta_{ij} \delta \left(\tau \right),
	\label{mtcfilt}
\end{eqnarray}

where $h_{ij}\left( \tau \right)$ is the impulse response of the filter between the
\emph{j}th input and the \emph{i}th output. Assuming that the number of input
channels $N$ is the same as the number of output channels $M$, the multichannel
coloring filter can be represented by a square matrix. 
The cross-spectral matrix of the output process can be obtained by the combination of
the cross-spectral matrix of the input and the frequency response of the filter:

\begin{equation}
	\mathbf{S}\left( \omega \right) = \mathbf{H}\left( \omega \right) \cdot \mathbf{I} \cdot \mathbf{H}^\dagger  \left( \omega \right).
	\label{eqn:S=HIH}
\end{equation}

Here $\mathbf{I}$ is the unit matrix corresponding to the cross-spectral matrix of
the input multichannel white noise process $\bm{\varepsilon} \left( t \right)$ and
$\mathbf{H}\left( \omega \right)$ is the frequency response matrix of the
multichannel filter. The problem of the generation of a multichannel noise series
with the given cross-spectral matrix starts from the identification of the noise
coloring filter $\mathbf{H}\left( \omega \right)$.

The eigendecomposition of the cross-spectral matrix $\mathbf{S}\left( \omega
\right)$ is defined as:

\begin{equation}
	\mathbf{S}\left( \omega \right) = \mathbf{V}\left( \omega \right) \cdot \bm{\Sigma} \left( \omega \right) \cdot \mathbf{V}^{ - 1} \left( \omega \right),
	\label{eigendecS}
\end{equation}

where $\mathbf{V}\left( \omega \right)$ and $\bm{\Sigma} \left( \omega \right)$ are
the eigenvector and eigenvalue matrices of the cross-spectral matrix
$\mathbf{S}\left( \omega \right)$. Since $\mathbf{S}\left( \omega \right)$ is
Hermitian, its eigenvector matrix is unitary, i.e. $\mathbf{V}\left( \omega
\right)\mathbf{V}^\dag \left( \omega \right) = \mathbf{I}$. Therefore, combining
equations (\ref{eqn:S=HIH}) and (\ref{eigendecS}), the noise coloring filter can be
obtained:

\begin{equation}
	\mathbf{H}\left( \omega \right) = \mathbf{V}\left( \omega \right) \cdot \sqrt{ \bm{\Sigma} \left( \omega \right)}. 
	\label{H=V*sqrt(D)}
\end{equation}

As $\bm{\Sigma} \left( \omega \right)$ is a diagonal matrix, $\sqrt {\bm{\Sigma}
\left( \omega \right)}$ is a diagonal matrix with elements given by the square root
of the elements of $\bm{\Sigma} \left( \omega \right)$.

\subsection{System Discretization}\label{subs:discretization}

Once the frequency response of the coloring filter $\mathbf{H}\left( \omega \right)$
is known, a discrete multichannel filter is required for the generation of discrete
synthetic noise data series. Discrete filters can be estimated by a least square fit
procedure carried out in the frequency domain. Such a process can produce a set of
discrete autoregressive moving average (ARMA) filters which together reproduce the
multichannel system frequency response to the given accuracy. The fitting process
is based on a modified version of the vector fitting algorithm
\cite{vectfit1,vectfit2} adapted to work in Z-domain \cite{vectfit3}. This procedure
allows the frequency response of the coloring filter to be fit with ARMA functions
expanded in partial fractions:

\begin{eqnarray}
	 \mathbf{H}\left( \omega \right) & = & \left( {\begin{array}{*{20}c}
	   {h_{11} \left( \omega \right)} & \cdots & {h_{1M} \left( \omega \right)}  \\
	   \vdots & \ddots & \vdots \\
	   {h_{M1} \left( \omega \right)} & \cdots & {h_{MM} \left( \omega \right)}  \\
	\end{array}} \right) \to \mathbf{H}\left( z \right) = \left( {\begin{array}{*{20}c}
	   {h_{11} \left( z \right)} & \cdots & {h_{1M} \left( z \right)}  \\
	   \vdots & \ddots & \vdots \\
	   {h_{M1} \left( z \right)} & \cdots & {h_{MM} \left( z \right)}  \\
	\end{array}} \right) \nonumber \\
	 h_{ij} \left( z \right) & = & \sum\limits_{k = 1}^N {\frac{{r_{ij,k} }}{{1 - p_{ij,k} z^{ - 1} }}}.
\label{Hfromftoz}
\end{eqnarray}

The number of poles required to obtain a satisfactory fit of the model transfer
function is automatically determined by an iterative procedure in which the number
of poles is increased by one at each step of the fit loop. The iteration stops when
the mean square error between fit function response and model response comes to a
value smaller than the user-defined threshold. Since the fit is performed in
the Z-domain, the noise generation procedure turn out to be free from aliasing as
discussed in section \ref{sec:principles}.

It is worth noting that the eigenvectors in equation (\ref{eigendecS}) are defined
up to an arbitrary phase factor. This means that the columns of the
$\mathbf{V}\left( \omega \right)$ matrix can be multiplied by an arbitrary phase
factors $e^{i\phi }$ without changing their property of being eigenvectors of the
cross-spectral matrix. Such phase arbitrariness does not extend to the single
elements in the columns of $\mathbf{V}\left( \omega \right)$; they are elements of
the same eigenvector of $\mathbf{V}\left( \omega \right)$ and their phase relation
must be carefully preserved during the fit process because it is connected to the
correlation properties of the multichannel system. It can happen, after the
eigen-decomposition process, that the phase of the elements of $\mathbf{V}\left(
\omega \right)$ is such that the frequency response of the elements of
$\mathbf{H}\left( \omega \right)$ cannot be fit with stable poles. In that case, the
poles must be stabilized after the fit process by the application of an all-pass
filter \cite{papoulis}. The all-pass function substitutes unstable poles with the
inverse of their conjugates which are stable. Its magnitude (absolute value) is 1 at
each frequency. As already mentioned, the phase relation between the elements of
each column of $\mathbf{H}\left( z \right)$ must be kept constant to prevent the
corruption of the system correlation properties. The proper all-pass filter for the
elements of $\mathbf{H}\left( z \right)$ stabilizes the unstable poles of the given
$h_{ij} \left( z \right)$ and at the same time adds an extra phase in order to keep
the phase relation between the columns of $\mathbf{H}\left( z \right)$ constant.
Therefore each element of $\mathbf{H}\left( z \right)$ is modified according to:

\begin{eqnarray}
 h_{\alpha \beta } \left( z \right) & \to & h_{\alpha \beta } \left( z \right)
 \prod\limits_{k = 1}^{N_{\alpha \beta }^u } {\left( {\frac{{z - p_{\alpha \beta ,k} }}{{zp_{\alpha \beta ,k}^ *   - 1}}} \right)}
 \left\{ \prod\limits_{\gamma \neq \alpha} \left[
 \prod\limits_{h = 1}^{N_{\gamma \beta }^u } {\left( {\frac{{z - p_{\gamma \beta ,h} }}{{zp_{\gamma \beta ,h}^ *   - 1}}} \right)} \right] \right\} \nonumber \\
 \alpha ,\beta ,\gamma  & = & 1,\cdots, M.
 \label{allpass2}
\end{eqnarray}

Here, $N_{\alpha \beta}^u$ is the number of unstable poles in $h_{\alpha \beta}
\left( z \right)$. The function $\prod\limits_{k = 1}^{N_{\alpha \beta }^u } {\left(
{\frac{{z - p_{\alpha \beta ,k} }}{{zp_{\alpha \beta ,k}^ * - 1}}} \right)}$ has the
purpose of poles stabilization for the element $h_{\alpha \beta } \left( z \right)$
of $\mathbf{H}\left( z \right)$. The product $ \prod\limits_{\gamma \neq \alpha}
\left[ \prod\limits_{h = 1}^{N_{\gamma \beta }^u } {\left( {\frac{{z - p_{\gamma
\beta ,h} }}{{zp_{\gamma \beta ,h}^ * - 1}}} \right)} \right]$ provides an extra
phase coming from the poles stabilization procedure for the other elements of the
same column of $\mathbf{H}\left( z \right)$. In this way, each element of the matrix
$\mathbf{H}\left( z \right)$ comes with such a phase allowing the representation
with just stable poles and, at the same time, the original phase relation between
elements of the same column of $\mathbf{H}\left( z \right)$ is preserved. A second
fit step with stable poles provides usable filters for noise generation.

\subsection{Filter Initialization}\label{subs:initialization}

The output of a linear discrete causal filter is a function of the present and all
previous input values.
Since the input to the filter must start at some time $t = 0$, the output process
will consist of an unwanted filter transient response added to the desired
stationary random process. One possible approach to handle the problem of the filter
transients is to wait for the time necessary for the transients to decay to an
acceptable level. However the transient response is proportional to the filter
impulse response and if there are poles too near the unitary circle of the complex
plane the transient response could last for an unacceptably long time. Moreover the
practice of hand removing initial data is always inaccurate and can hide important
features or introduce fake signals especially when the spectrum spans several
decades in frequency. Therefore the filter for the noise generation should be always
properly initialized.

\begin{figure}
	\centering
		\includegraphics[width=0.7\textwidth]{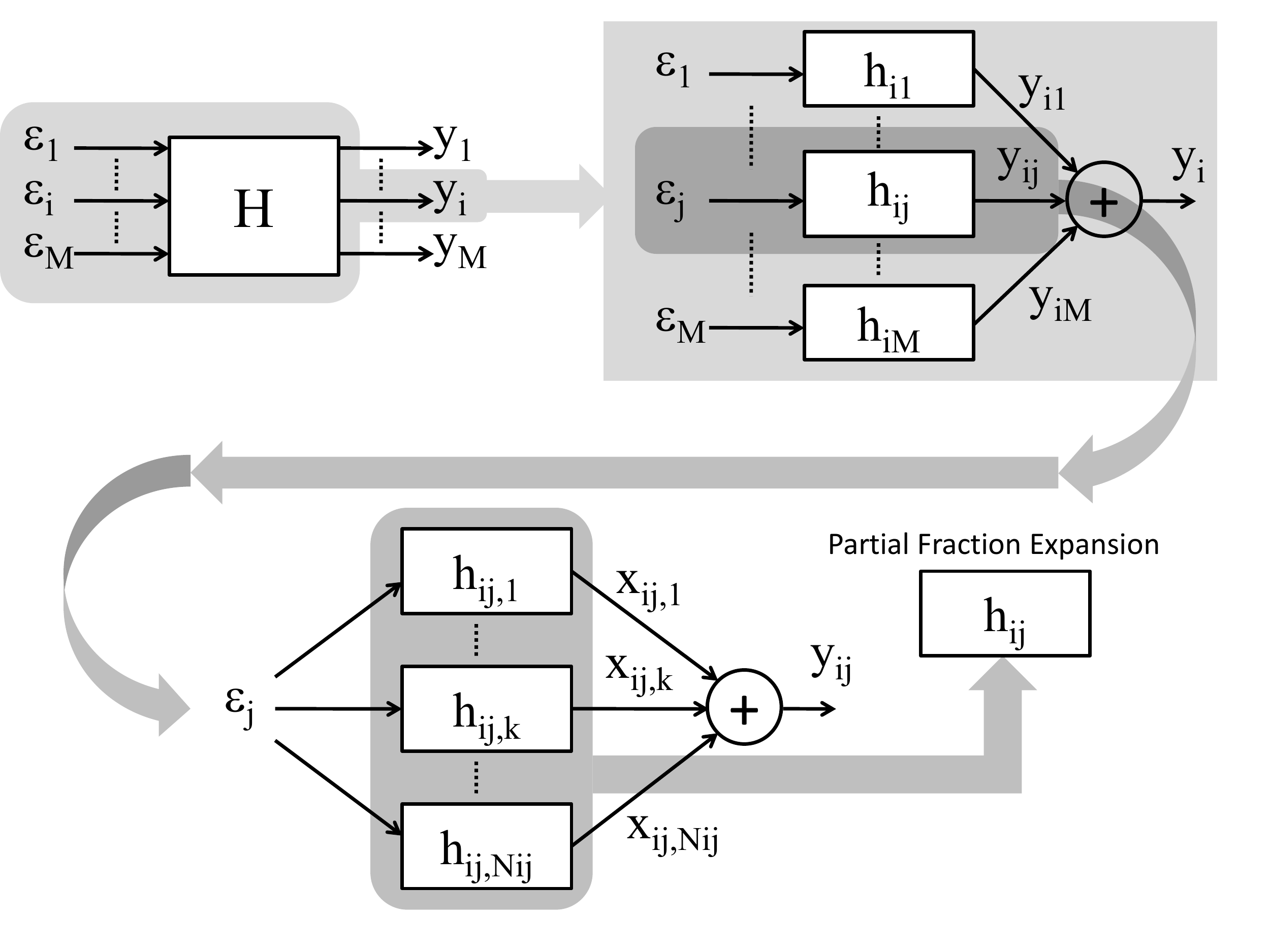}
	\caption{Scheme for the application of a multichannel filter expanded in partial fractions.}
	\label{fig:Figure1}
\end{figure}

We are searching for an initialization for the recursive equation of a ARMA process
written as sum of partial fractions (Figure (\ref{fig:Figure1}), Equation
(\ref{Hfromftoz})):

\begin{eqnarray}
 x_{ij,k} \left( n \right) & = & p_{ij,k} x_{ij,k} \left( {n - 1} \right) + r_{ij,k} \varepsilon_j \left( n \right) \nonumber \\
 y_{ij} \left( n \right) & = & \sum\limits_{k = 1}^{N_{ij} } {x_{ij,k} \left( n \right)}.
 \label{recurrence1}
\end{eqnarray}

The recursive equation (\ref{recurrence1}) calculates the output of the system at a
given time on the basis of the input $\varepsilon_j \left( n \right)$ at the same
time and the information from the previous output $x_{ij,k} \left( {n - 1} \right)$.
In the present case the input process $\varepsilon_j \left( n \right)$ is an element
of a discrete multichannel unitary variance white noise process such that:

\begin{equation}
\left\langle {\varepsilon _i \left( n \right)\varepsilon _j \left( m \right)} \right\rangle  = \delta _{i,j} \delta _{n,m}.
\label{deltacorrelation}
\end{equation}

In order to properly initialize the recursive equations for the implementation of
the multichannel filter, it is necessary to calculate the covariance matrix of the
initial states $x_{ij,k} \left( 0 \right)$. It should be considered that the $M^2$
processes represented in equation (\ref{mtcfilt}) are not independent from each
other. A combined process should be defined in which all the recursive equations
(equation \ref{recurrence1}) for each of the $M^2$ processes are incorporated.
Readily it is seen that, thanks to the delta correlation properties
(\ref{deltacorrelation}) of the input signals, the processes applied to different
input data series are independent. This means that, instead of building a single
combined process, one has to build $M$ independent processes which combine those
applied to the same input. The new processes can then be written as:

\begin{eqnarray}
 \bm{\chi} _j \left( n \right) & \to & \left\{ \begin{array}{l}
 x_{1j,1} \left( n \right) = p_{1j,1} x_{1j,1} \left( {n - 1} \right) + r_{1j,1} \varepsilon _j \left( n \right) \\ 
  \vdots  \\ 
 x_{1j,N_{1j} } \left( n \right) = p_{1j,N_{1j} } x_{1j,N_{1j} } \left( {n - 1} \right) + r_{1j,N_{1j} } \varepsilon _j \left( n \right) \\
 \vdots \\ 
 x_{Mj,1} \left( n \right) = p_{Mj,1} x_{Mj,1} \left( {n - 1} \right) + r_{Mj,1} \varepsilon _j \left( n \right) \\ 
  \vdots  \\ 
 x_{Mj,N_{Mj} } \left( n \right) = p_{Mj,N_{Mj} } x_{Mj,N_{Mj} } \left( {n - 1} \right) + r_{Mj,N_{Mj} } \varepsilon _j \left( n \right) \\ 
 \end{array} \right. \nonumber \\
 j & = & 1, \cdots, M.
 \label{combinedprocess}
\end{eqnarray}

The covariance of such processes (\ref{combinedprocess}) can be written as:

\begin{eqnarray}
 \left\langle {\chi _{i,\alpha } \left( n \right)\chi _{j,\beta }^ *  \left( m \right)} \right\rangle  & = & \left\langle {p_{i,\alpha } \chi _{i,\alpha } \left( {n - 1} \right)p_{i,\alpha }^ *  \chi _{j,\beta }^ *  \left( {m - 1} \right)} \right\rangle  + \left\langle {r_{i,\alpha } \varepsilon _i \left( n \right)r_{i,\beta }^ *  \varepsilon _j^ *  \left( m \right)} \right\rangle  \nonumber \\
 \alpha  & = & 1, \ldots ,N_{1i} + \cdots + N_{Mi}  \nonumber \\ 
 \beta  & = & 1, \ldots ,N_{1j}  + \cdots + N_{Mj}  \nonumber \\
 i,j & = & 1, \cdots, M.
 \label{cov1}
\end{eqnarray}

Where again, the symbol $\left\langle \right\rangle$ represents the expectation value operator. Assuming stationary processes:

\begin{equation}\label{cov2}
\left\langle {\chi _{i,\alpha } \left( n \right)\chi _{j,\beta }^ *  \left( m \right)} \right\rangle  = R_{ij,\alpha \beta } \left( {n,m} \right) = R_{ij,\alpha \beta } \left( {n - m} \right),
\end{equation}

and thanks to the properties of the input functions (\ref{deltacorrelation}):

\begin{equation}\label{cov3}
R_{ij,\alpha \beta } \left( {n - m} \right) = \frac{{r_{i,\alpha } r_{j,\beta }^ *  }}{{1 - p_{i,\alpha } p_{j,\beta }^ *  }}\delta _{ij} \delta \left( {n - m} \right),
\end{equation}

which provides the desired covariance for the first state of the recurrence sequences:

\begin{equation}\label{cov4}
R_{j,\alpha \beta } \left( 0 \right) = \frac{{r_{j,\alpha } r_{j,\beta }^* }}{{1 - p_{j,\alpha } p_{j,\beta }^* }}.
\end{equation}

Initial states for the filter recurrence sequence can then be generated by a
multivariate noise generator according to the given covariance (\ref{cov4}). As an
alternative, they can be calculated from random independent variables through a
linear transformation \cite{stein} of the type:

\begin{equation}\label{initstate1}
\bm{\chi} _j \left( 0 \right) = \mathbf{A}_j  \cdot \bm{\eta} _j,
\end{equation}

where $\bm{\eta} _j$ is a column vector of $N_{1j} + \cdots + N_{Mj}$ independent
zero mean and unit variance random numbers and ${\mathbf{A}}_j$ is a $\left( {N_{1j}
+ \cdots + N_{Mj} } \right)\times\left( {N_{1j} + \cdots + N_{Mj} } \right)$
transformation matrix. If $R_{j,\alpha \beta } \left( 0 \right)$ is calculated for
the variables in equation (\ref{initstate1}), it is found:

\begin{eqnarray}\label{initstate2}
\mathbf{R}_j \left( 0 \right) & = & \left( {{\mathbf{A}}_j \cdot \bm{\eta} _j } \right)\cdot\left( {{\mathbf{A}}_j \cdot \bm{\eta} _j } \right)^\dagger \nonumber \\
 & = & {\mathbf{A}}_j \cdot{\mathbf{I}}\cdot{\mathbf{A}}_j ^\dagger,
\end{eqnarray}

and it is readily seen that:

\begin{equation}\label{initstate3}
{\mathbf{A}}_j  = \mathbf{V}_j  \cdot \sqrt {\bm{\Sigma} _j },
\end{equation}

where $\mathbf{V}_j$ and $\bm{\Sigma} _j$ are the eigenvector and eigenvalue
matrices of $\mathbf{R}_j \left( 0 \right)$.

\section{A case study, LTP along $X$ axis}

\subsection{Response model and fit}

An application of the noise generation procedure is presented for a two channel
system simulating the LISA Technology Package (LTP) along the principal measurement
axis \cite{Armano, Anza, McNamara, Bortoluzzi1, Bortoluzzi2, LISA7Anneke,
LISA7Martin, LISA7Luigi}. The complete set of algorithms are available as MATLAB
tools in the framework of the LTPDA toolbox \cite{LISA7Martin, matlab} and can be
freely downloaded, together with the complete toolbox, at the LTPDA project web page
\cite{ltpda}.

\begin{figure}[htbp]
	\centering
		\includegraphics[width=0.7\textwidth]{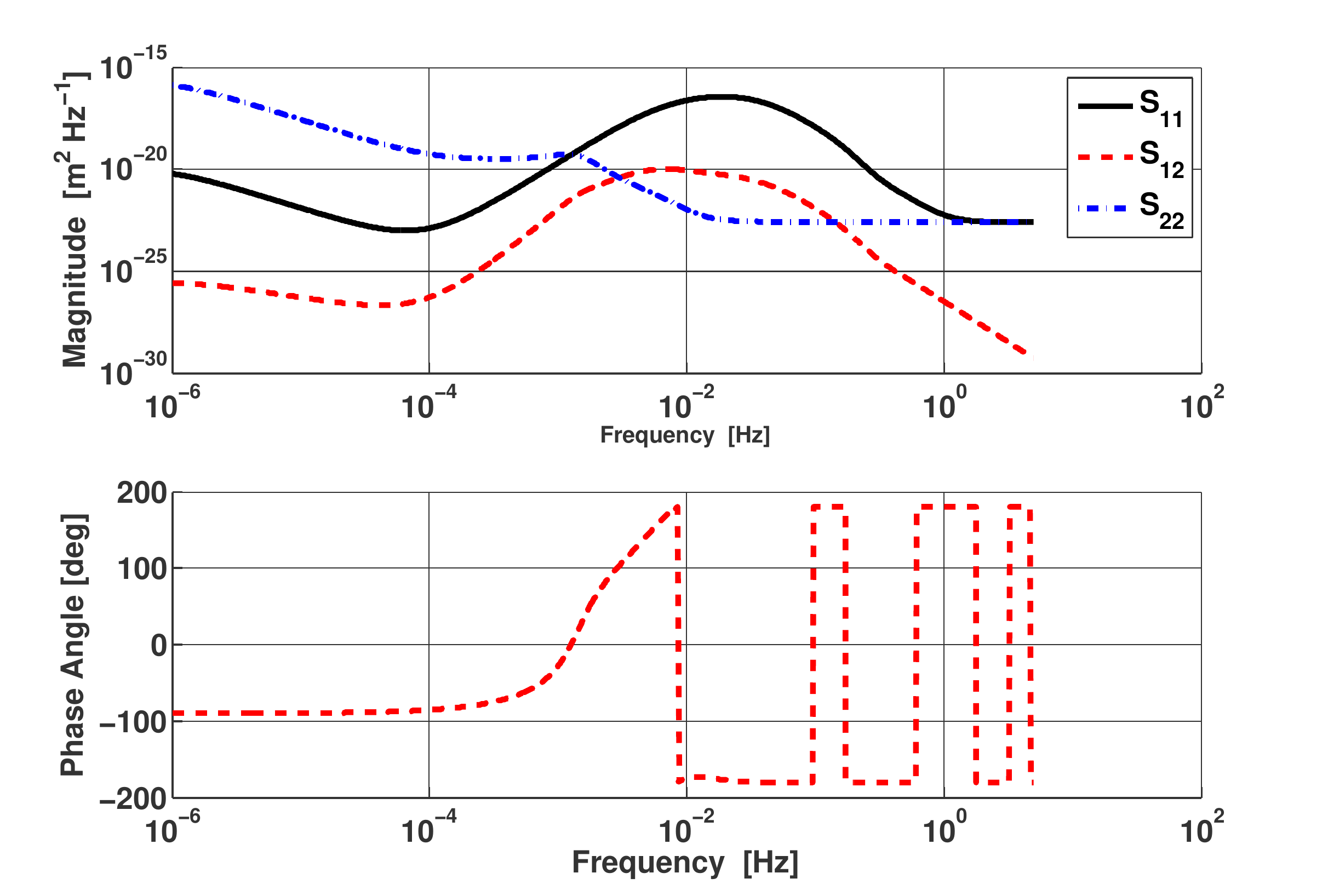}
	\caption{Model power spectra and cross-spectrum for the signals at the output of the two channels. $S_{11}$ and $S_{22}$ are real values so they do not appear in the bottom plot.}
	\label{fig:Figure2}
\end{figure}

\begin{figure}[htbp]
  \centering
  \subfloat[]{\label{fig:Figure3a}\includegraphics[width=0.5\textwidth]{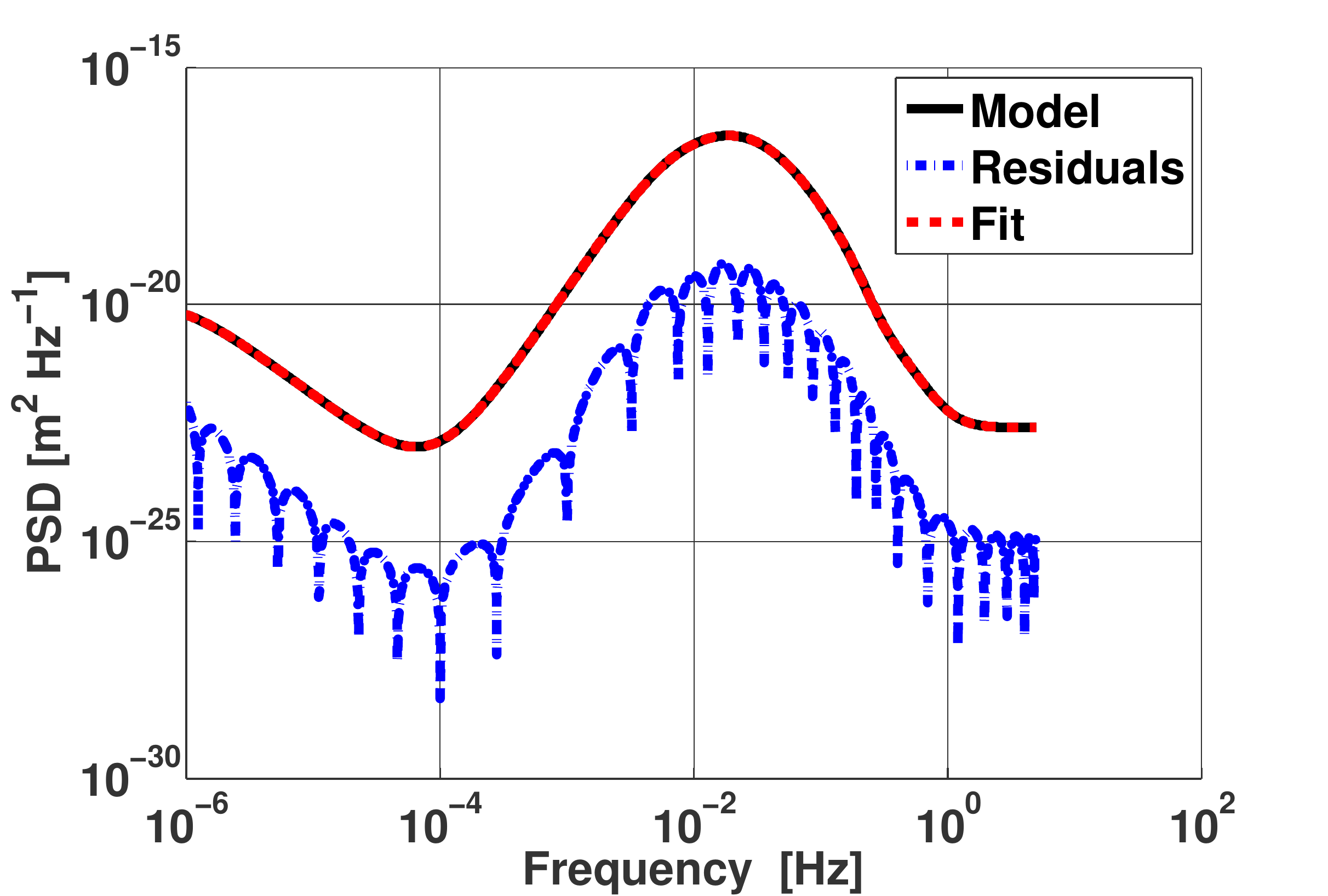}}                
  \subfloat[]{\label{fig:Figure3b}\includegraphics[width=0.5\textwidth]{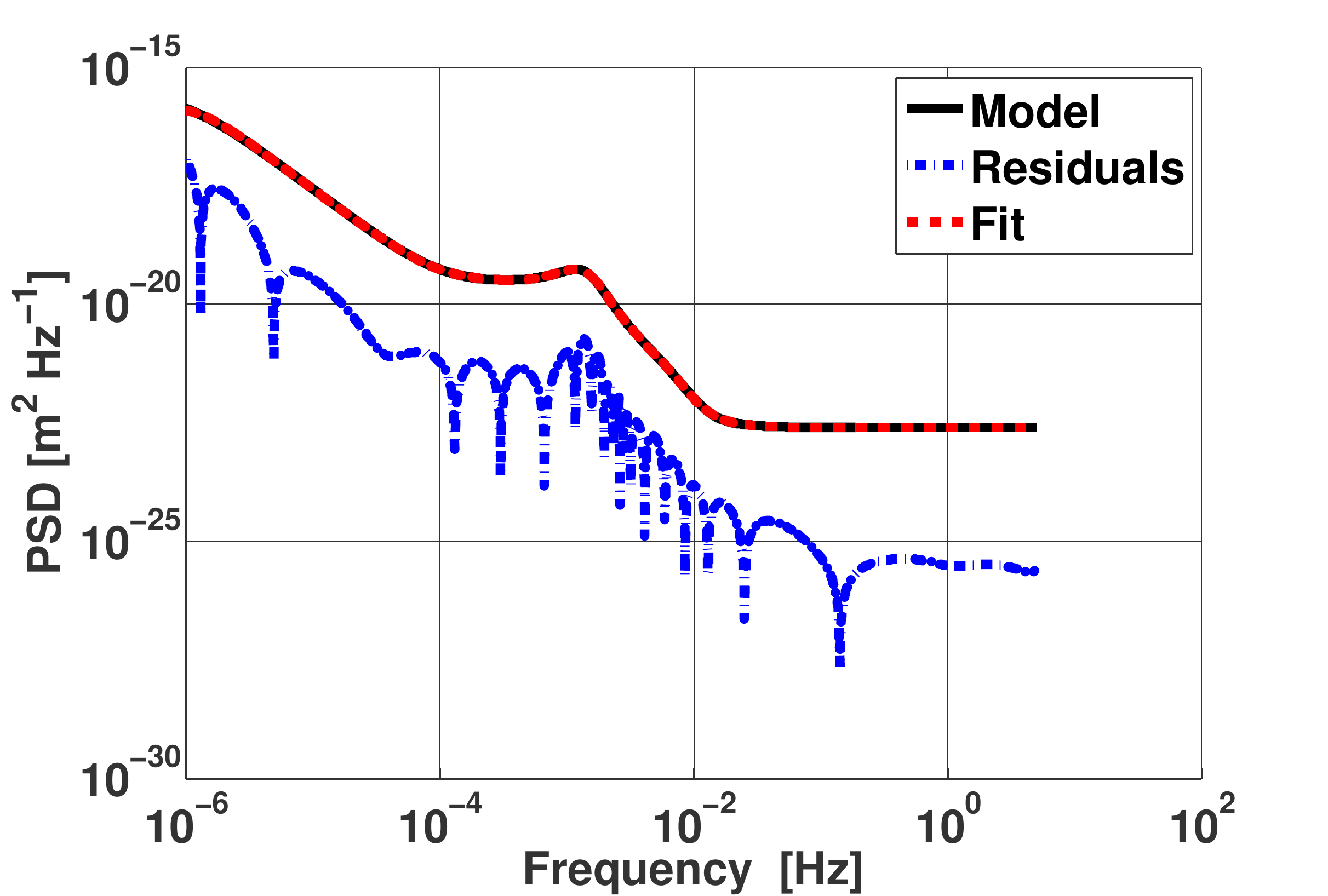}}
  \caption{Comparison between model power spectral densities and fit result. a) Output of the first channel. b) Output of the second channel.}
  \label{fig:Figure3}
\end{figure}

The expected power spectra and cross-power spectrum at the output of the system can
be calculated (Figure (\ref{fig:Figure2})) on the basis of some assumptions on the
properties of input noise sources \cite{Bortoluzzi1}. The noise coloring filters can
be calculated following the procedure described in paragraphs
\ref{subs:noisecoloring} and \ref{subs:discretization}. A frequency domain fit is
performed on the models for the coloring filters obtained by eigendecomposition
(frequency by frequency) of the expected cross-spectral density matrix. The fit
procedure takes around $200$ seconds on a standard desktop machine~\footnote{For the
example presented here, a $32$ bit Windows machine equipped with $4$ GB RAM and an
Intel core DUO $2.26$ GHz processor was used.}. The four transfer functions $h_{11}
\left(z\right)$, $h_{12} \left(z\right)$, $h_{21} \left(z\right)$ and $h_{22}
\left(z\right)$ have respectively $25$, $30$, $28$ and $30$ poles. The fit loop
stops when the mean square error between fit function response and model response is
smaller than $1 \times 10^{-4}$. The response of the filter designed to reproduce
the cross-spectral density can be calculated according to equation
(\ref{eqn:S=HIH}). It can then be compared with the model cross-spectral density
of the system as reported in figure \ref{fig:Figure3}.

\begin{figure}[htbp]
  \centering
  \subfloat[]{\label{fig:Figure4a}\includegraphics[width=0.5\textwidth]{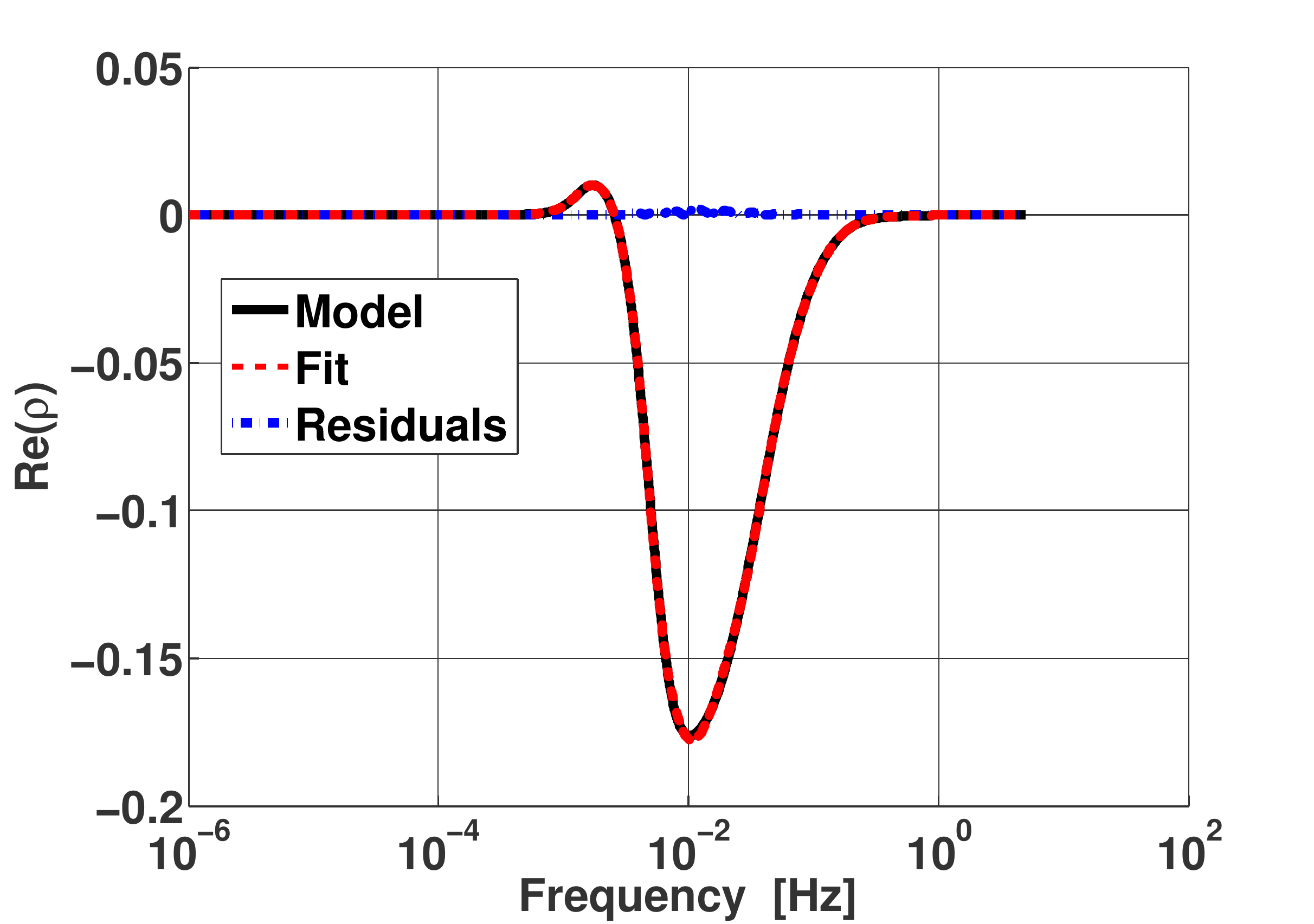}}                
  \subfloat[]{\label{fig:Figure4b}\includegraphics[width=0.5\textwidth]{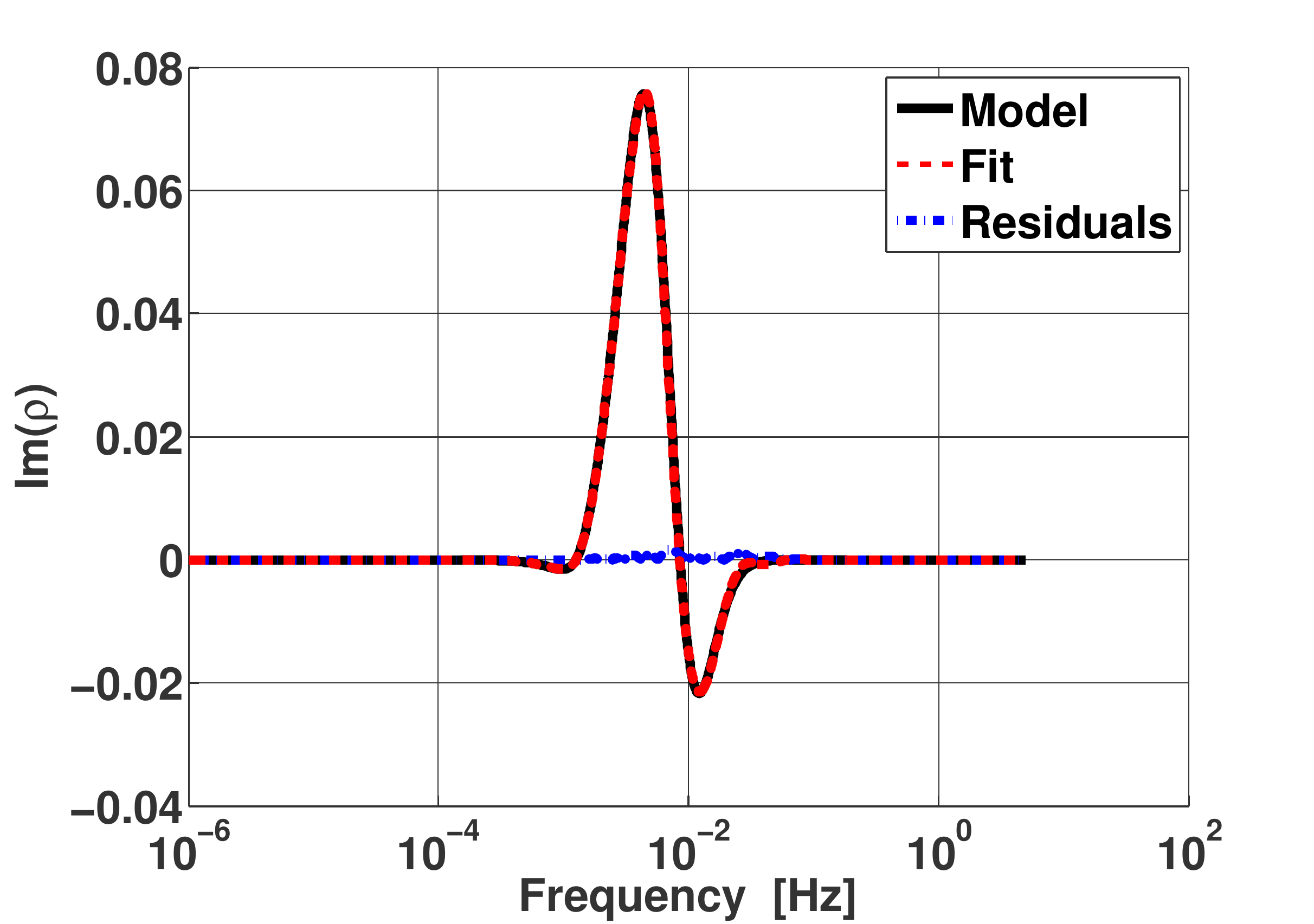}}
  \caption{Comparison between expected coherence and fit result. a) Real part. b) Imaginary part.}
  \label{fig:Figure4}
\end{figure}

In order to compare the correlation properties of expected model with the fit
results, it is useful to introduce the complex cross-coherence:

\begin{equation}\label{eqn:coh}
\rho\left( \omega \right)  = \frac{{S_{12} \left( \omega \right) }}{{\sqrt {S_{11} \left( \omega \right) S_{22} \left( \omega \right) } }},
\end{equation}

where $S_{12} \left( \omega \right)$, $S_{11} \left( \omega \right)$ and $S_{22}
\left( \omega \right)$ are cross-spectrum and power spectra of the first and second
channels. The real and imaginary part of the cross-coherence for the expected and
fit cross-spectral matrices are reported in figure \ref{fig:Figure4}.

Any discrepancy between the expected model and the fit model can be considered as
a systematic error in the procedure, whose influence on the process can be minimized
by increasing the fit accuracy. Clearly this has a computational cost in terms of
the number of poles required to match the accuracy goal and on the amount of time
required to complete the fit loop. Hereafter, the fit model will be considered as
the reference model.

\subsection{Noise generation tests}

Once the two channel noise coloring filter is obtained it can be used to generate a
two channel noise data series according to the procedure described in paragraph
\ref{subs:initialization}. Data series are $3 \times 10^5$ seconds long at a
sampling rate of $10$ Hz. The chosen rate is the same as the LTP
experiment operations, e.g., the control forces acting on test masses and the
spacecraft will be calculated by controllers on the basis of $10$ Hz sampled data
streams.

\begin{figure}[htbp]
  \centering
  \includegraphics[width=0.7\textwidth]{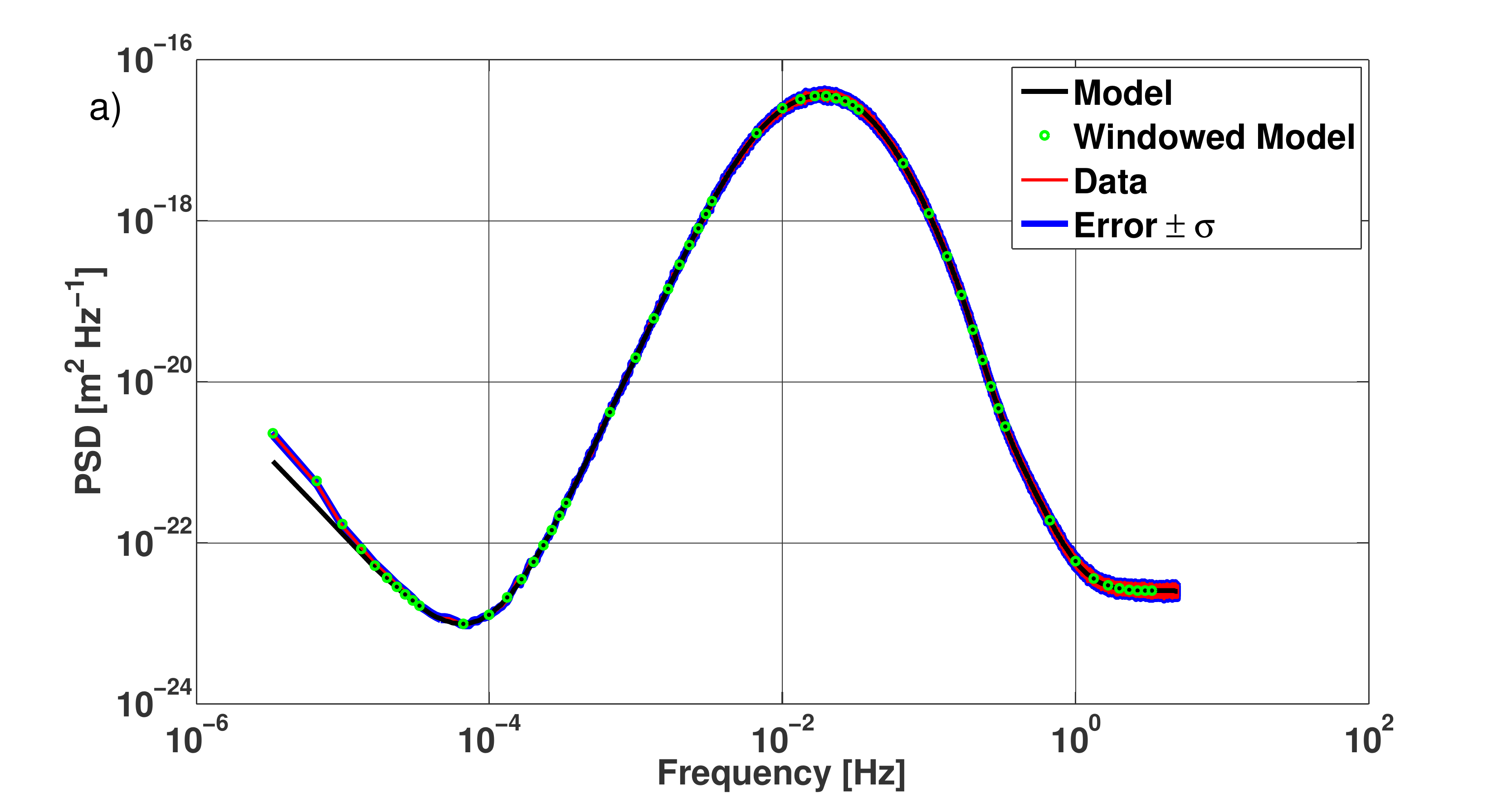}               
 	\includegraphics[width=0.7\textwidth]{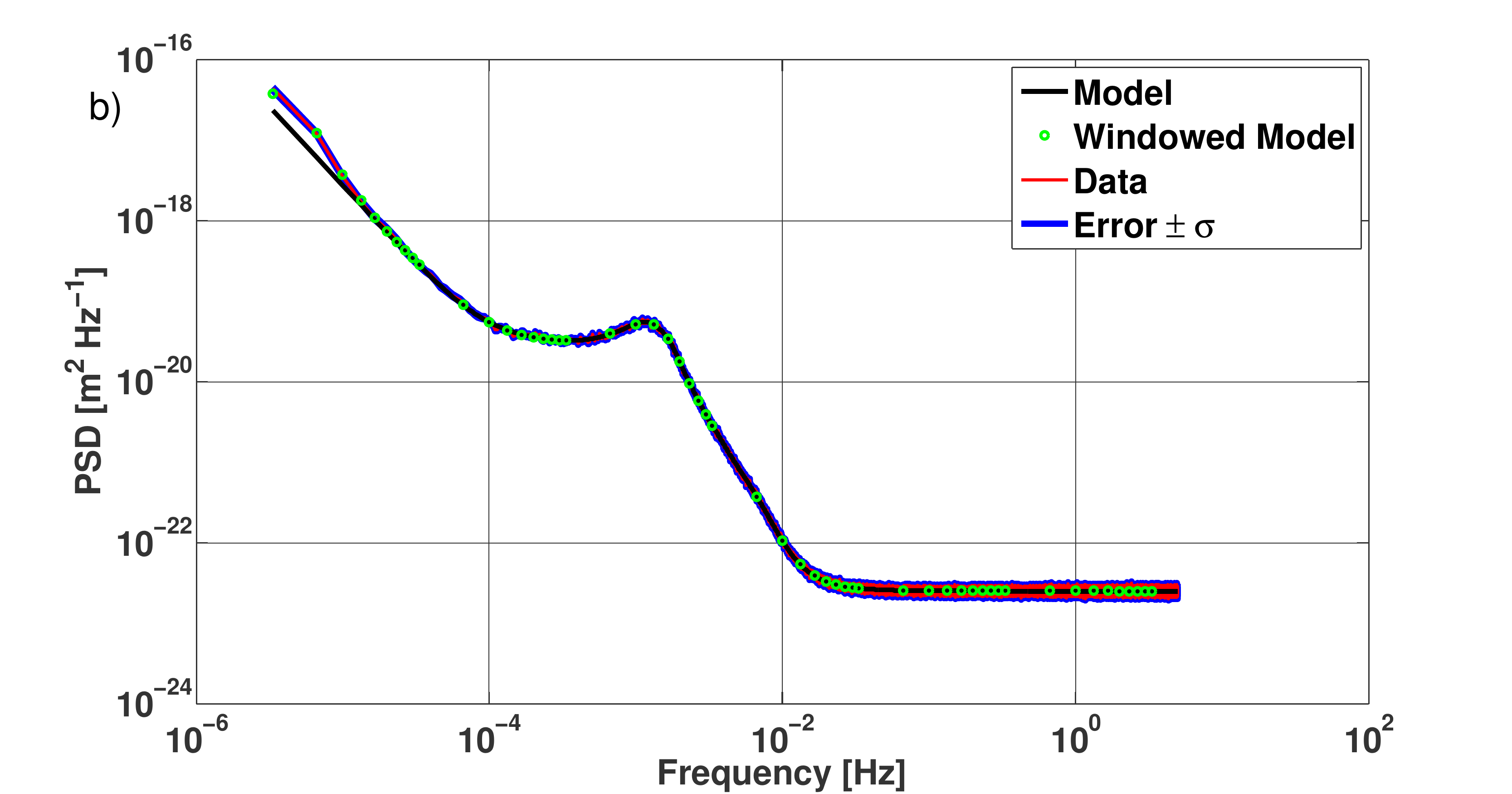}
  \caption{Averaged power spectral density compared with reference model. a) First channel. b) Second channel.}
  \label{fig:Figure5}
\end{figure}

In order to realize a statistically meaningful test, $N = 500$ independent
realizations of the two channel process were generated. Power spectra of the two
channels are calculated with the windowed periodogram method using a 4-term
Blackman-Harris window \cite{Harris}. The choice of such a window is justified by
the requirements in terms of spectral leakage performances. A 4-term
Blackman-Harris window, having the highest side-lobe level of -$92$ dB (relative to
the main lobe level) \cite{Harris}, is indeed one of the best-performing available
window in terms of spectral leakage suppression. The $N$ realizations of the power
spectrum were averaged and compared with the reference model expectation (figure
\ref{fig:Figure5}).

The reported uncertainty is calculated under the assumption that $\sigma_{mean}
\left( \omega \right) = \frac{\sigma_{pop} \left( \omega \right)}{\sqrt{N}}$, where
$\sigma_{pop} \left( \omega \right)$ is the sample standard deviation of the spectra
population at a given frequency. In doing this we have considered that, since the
power spectrum of a $\chi^2 _2$ distributed variable \cite{Jenkins}, the mean of $N$
independent realization of the same spectrum will also be $\chi^2 _{2N}$
distributed. Since in our case $N = 500$, the average of the spectra is $\chi^2$
distributed with $1000$ degrees of freedom, and such a variable can be considered to
be Gaussian distributed with reasonable accuracy \footnote{The maximum deviation
between the corresponding $\chi^2 _{1000}$ and the normal cumulative distributions
is $0.006$.}. The expected standard deviation (normalized to the mean) for
the equivalent normal distribution is $0.045$ where we measure on average
$\sigma_{mean} = 0.044 \pm 0.002$ \footnote{The sample spectrum is approximately
$\chi^2 _2$ distributed around the expected value $S$. Thus, in practice, the
distribution of the sample spectrum is $\frac{S}{2} \chi^2 _2$, whereas the
distribution of an average on $N$ realization is $\frac{S}{2N} \chi^2 _{2N}$. The
Gaussian distribution equivalent to a $\chi^2 _{2N}$ has expectation value of $2N$
and standard deviation $2\sqrt{N}$ therefore the Gaussian equivalent to a
$\frac{S}{2N} \chi^2 _{2N}$ distribution has an expectation value $S$ and standard
deviation $S \frac{\sqrt{N}}{N}$.}.

In the procedure for the calculation of the power spectrum, data are multiplied in
the time domain for the time response of the window function. As this operation
corresponds to a convolution in the frequency domain, the reference model must
include also the effect of the window function. Windowed spectra can be calculated
as:

\begin{equation}
	S_w\left(\psi = \frac{2 \pi k}{N}\right) = \frac{1}{2 \pi N} \int\limits_{-\pi}^{\pi} S\left(\Omega\right)\left|\sum_{q=0}^{N-1} w_q e^{\imath q \left(\Omega - k \frac{2 \pi}{N}\right)}\right|^2 \, d\Omega,
	\label{eqn:winspect}
\end{equation}

where $N$ is the number of samples in the data series and
$w_q$ are the time samples of the window function. The integral in equation
(\ref{eqn:winspect}) is numerically evaluated, and the results are reported in
figure \ref{fig:Figure5}. As can be seen, the effect of the window convolution is
visible at the lowest frequencies, were the departure from the reference model is
remarkable.

A quantitative analysis of the results is better performed with the introduction of the variable:

\begin{equation}\label{eqn:DS}
\Delta S \left( \omega \right) = \frac{S_{yy} \left( \omega \right) - S \left( \omega \right)}{S \left( \omega \right)}.
\end{equation}

$S \left( \omega \right)$ represents the expected value for the spectrum (at each
frequency) and $S_{yy} \left( \omega \right)$ is the estimated spectrum (averaged
over $500$ realizations) at the given frequency. $\Delta S$ can be considered
distributed in accordance to a $\frac{\chi^2_M - M}{M}$ function. Therefore its
expectation value is $0$.

\begin{figure}[htbp]
  \centering
  \includegraphics[width=0.7\textwidth]{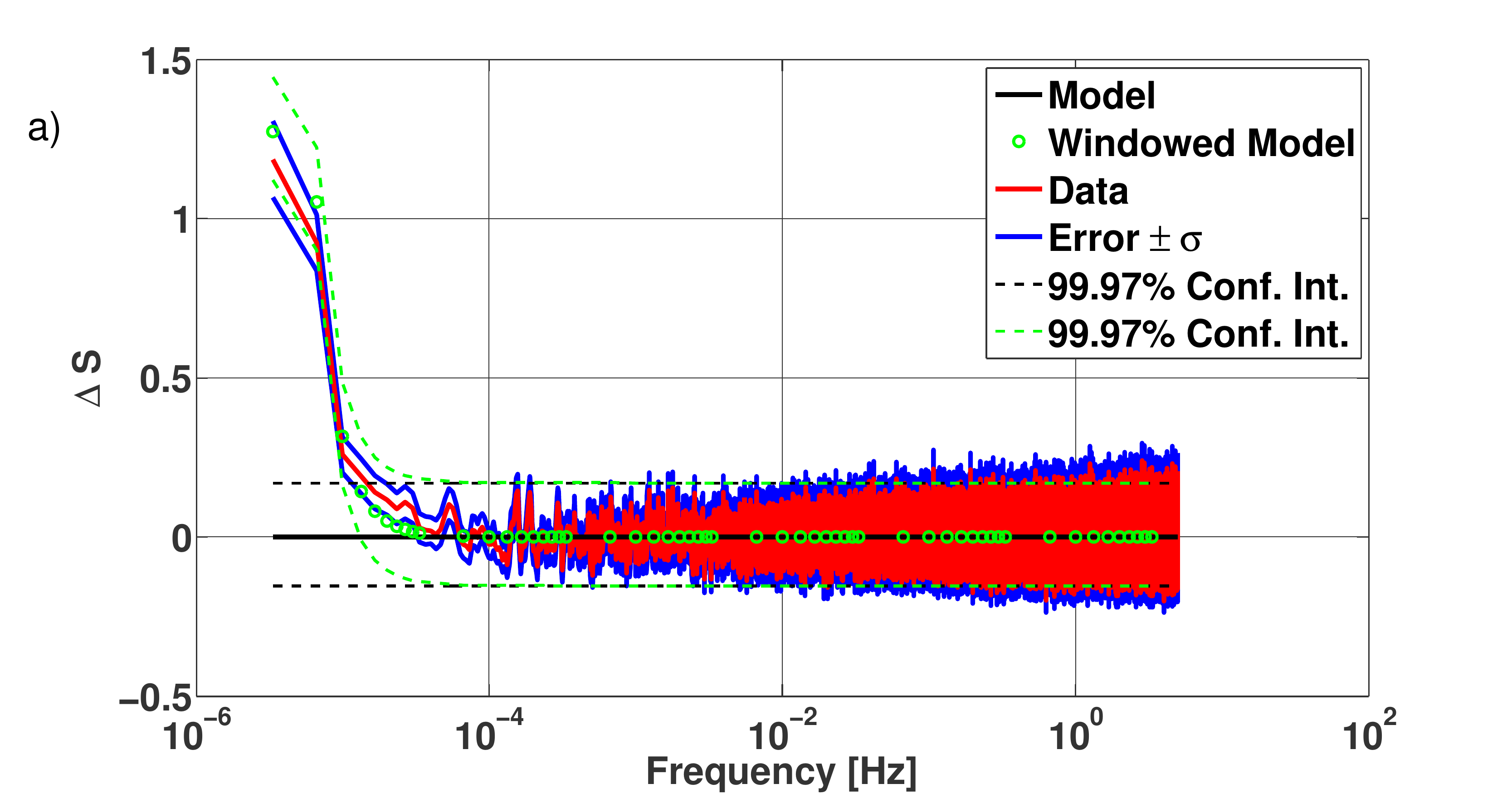}               
  \includegraphics[width=0.7\textwidth]{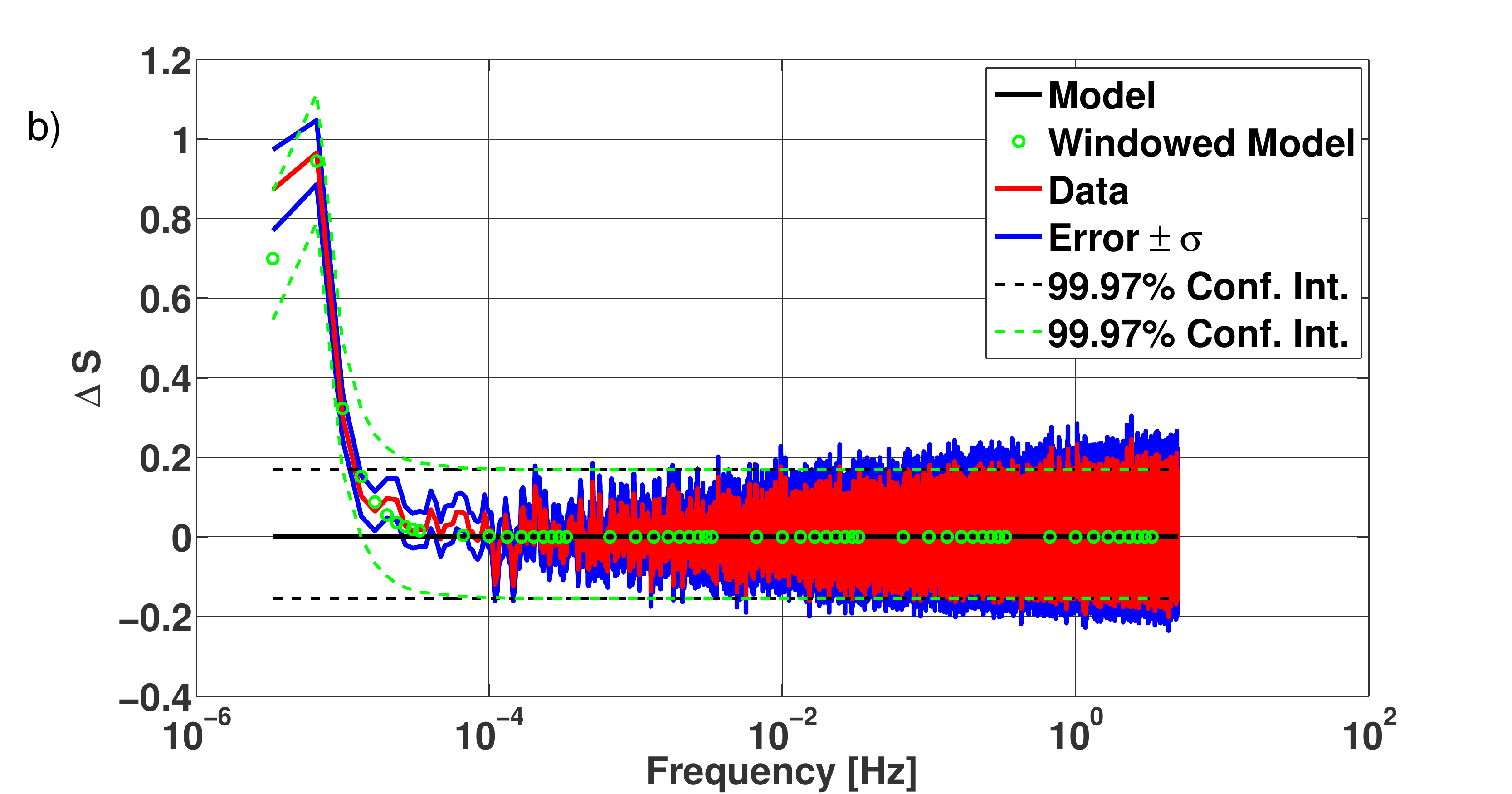}
  \caption{$\Delta S \left( \omega \right)$ calculated for a) First channel and b) Second channel. Simulated results are compared with the model expectation and the windowed model expectation. A $99.97 \%$ confidence interval is calculated for quantitative comparison with the models.}
  \label{fig:Figure6}
\end{figure}

$\Delta S$ is calculated for the simulated data and the windowed model with respect
to the reference model. Results are reported in figure \ref{fig:Figure6}. A
$99.97\%$ confidence interval is calculated on the basis of the statistical
properties of $\Delta S$.

The effect of the window on the spectra calculation is noticeably high on the first
$3$ frequency bins, where the deviation from the reference model exceeds the
confidence interval. In addition, it is clearly observable up to the $10^{th}$
bin. The windowed model and the simulated data are consistent on the basis of the
chosen confidence region.

A considerable number of data points, especially at high frequencies, lie outside
the confidence levels, and it is of fundamental importance to assess if such
outliers are caused by the random nature of the data, or if they come from
systematic errors in the spectral estimation or data generation processes. The
confidence levels at $99.97\%$ define a region in which the data are expected to lie
with that probability. This also means that in  $0.03\%$ of the observations an
outlier can be observed. As we are dealing with datasets of $1.5 \times 10^6$ points
the number of expected outliers is high.

\begin{figure}[htbp]
	\centering
		\includegraphics[width=0.7\textwidth]{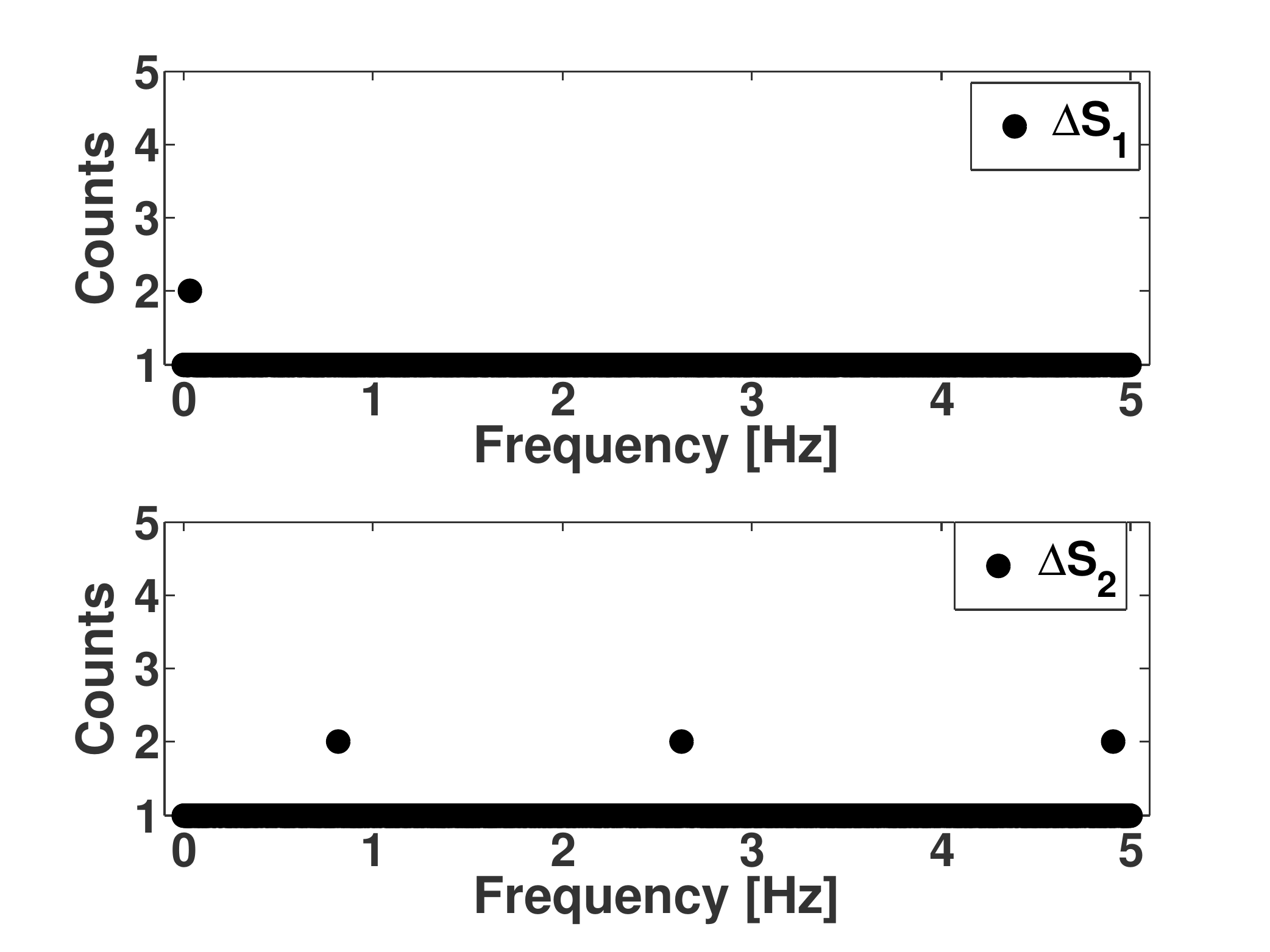}
	\caption{Histogram of the outliers frequencies for $\Delta S_1$ and $\Delta S_2$.}
	\label{fig:Figure7}
\end{figure}

In order to distinguish between systematic outliers and statistical outliers the
averaging process over $500$ independent realizations was repeated $5$ times and the
frequencies at which the values of $\Delta S$ were outside the defined confidence
interval were recorded. The first $10$ frequency bins are excluded from the
numbering because they are systematically affected by the spectral widow effect.
Figure \ref{fig:Figure7} reports a histogram of the cumulative count of the outliers
frequencies for the $5$ different realizations. If an outlier is originated by a systematic
error, then it is expected to be counted $5$ times. As can be observed from
figure \ref{fig:Figure7}, the maximum value obtained is $2$ for both channels; this
is a definitive indication of the statistical nature of the observed outliers.

\begin{figure}[htbp]
  \centering
  \subfloat[]{\label{fig:Figure8a}\includegraphics[width=0.5\textwidth]{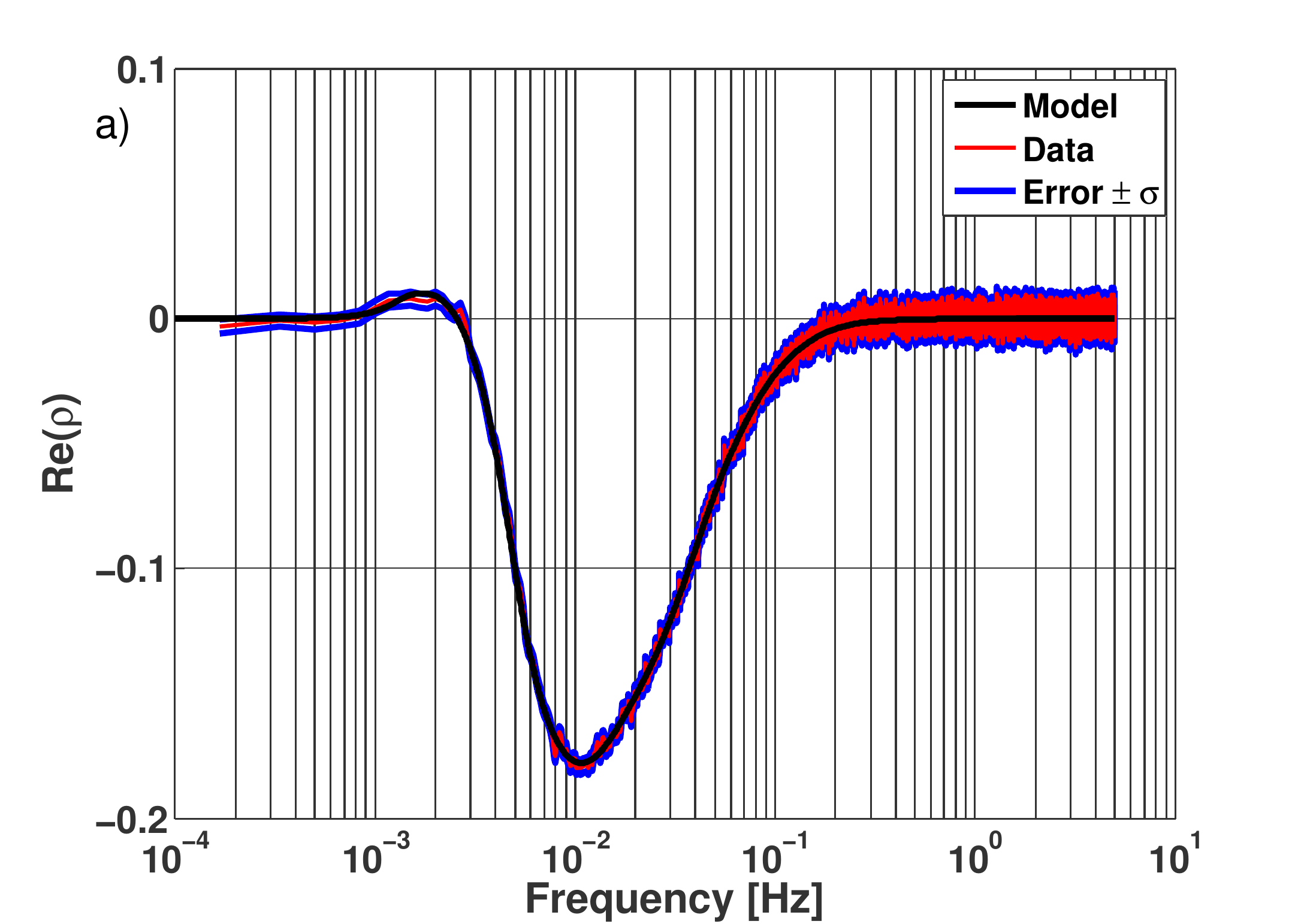}}                
  \subfloat[]{\label{fig:Figure8b}\includegraphics[width=0.5\textwidth]{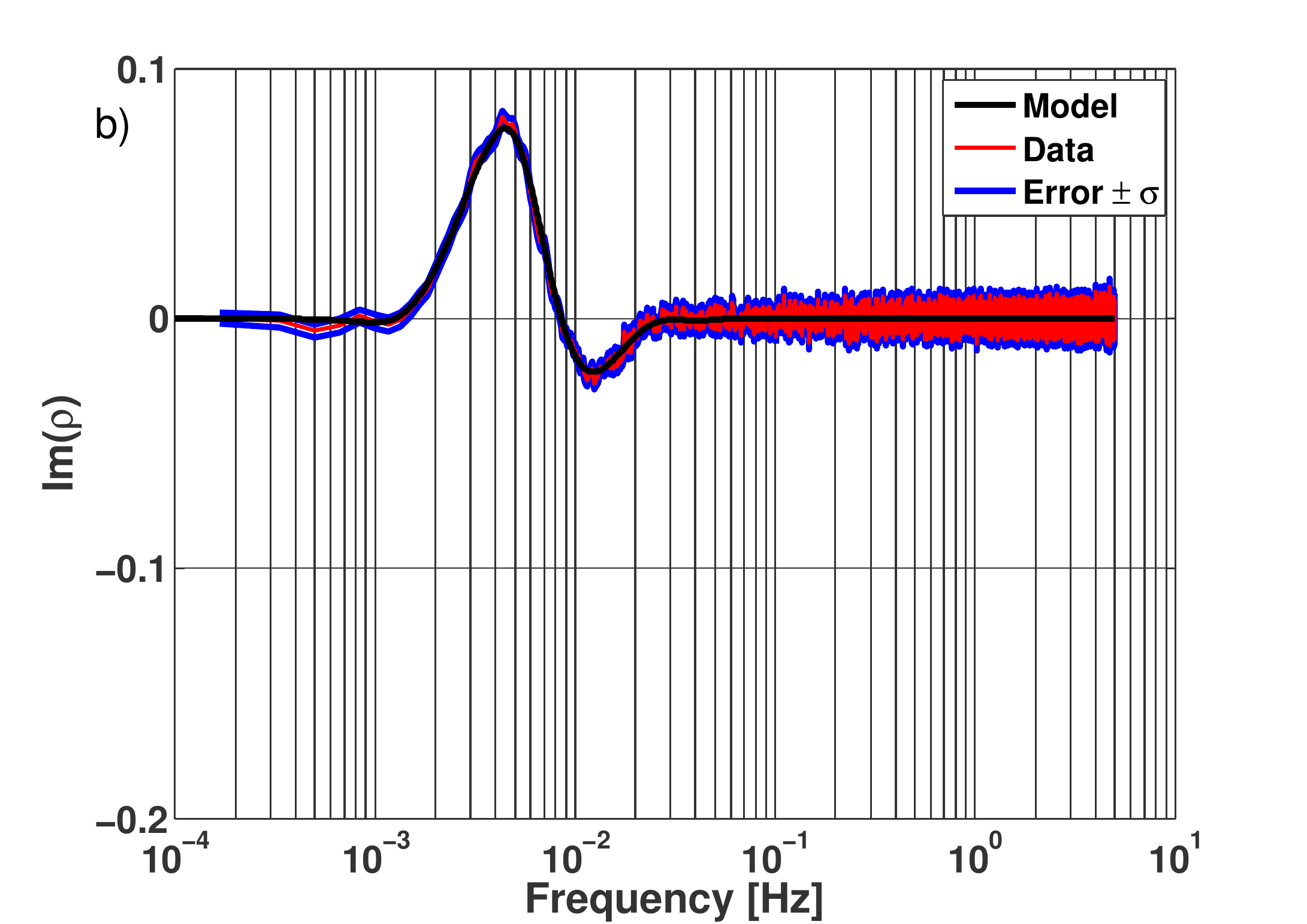}}
    \caption{Averaged cross-coherence compared with the reference model. a) Real part. b) Imaginary part.}
  \label{fig:Figure8}
\end{figure}

As stated above, the correlation properties of the data series can be explored with
the sample coherence calculated as in equation (\ref{eqn:coh}). Power spectra and
cross-spectrum were estimated with the averaged Welch periodogram method using a
4-term Blackman-Harris window over 145 data segments $6 \times 10^4$ points long.
The separate 500 realizations are then averaged and, assuming the averaged process
is approximately normally distributed, the error on the estimation is calculated as
described above. Results are reported in figure \ref{fig:Figure8} and compared with
the expectation from the reference model. Since the coherence is constructed from a
ratio between cross-spectrum and power spectra the effect of the window on the
lowermost frequency bins is strongly attenuated.

The simulated data (averaged over $500$ realizations) and the reference model are in
satisfactory agreement within the tolerance region defined by the uncertainty. On
the basis of the above discussion, the oscillations observed in the coherence curves
(figure \ref{fig:Figure8}) can be associated with the statistical fluctuations
caused by the random nature of the data.

\section{Conclusions}

A robust procedure for the generation of multichannel stationary noise with a given
cross-spectral matrix is reported. Based on some assumptions on the noise sources
acting on the system under study, an expected model for the cross-spectral matrix of
the multichannel output noise can be developed. From such a model the noise
coloring filters are identified by an eigendecomposition of the cross-spectral
matrix (frequency by frequency) and a frequency domain fit procedure. A
multichannel colored noise data series can then be generated from a multichannel
$\delta$ correlated random noise process provided that the recurrence equations are
properly initialized in order to avoid transients at the beginning of the noise
sequence. It is demonstrated that the only source of systematic errors in the
process is associated with the fit procedure. On the other hand, the accuracy of the
fit can be increased at the expense of the computational cost of the whole process;
this, in principle, ensures that the process reaches the desired accuracy. An
average over $500$ independent realizations of the multichannel noise process has
demonstrated the statistical consistency between generated noise and the reference
model if the effect of the spectral window is taken into account. Oscillations in
the averaged spectra with respect to the model can be unambiguously attributed to
statistical fluctuations. The analysis reported demonstrates that the tool can be
applied for the calibration of spectral estimators in experiments where noise
spectral energy content must be estimated with very high accuracy, as is the case
for the LTP experiment. MATLAB based algorithms are available for free download at
the LTPDA project web page.

\end{document}